\documentclass[twocolumn,aps,amsmath,amssymb,prd,showpacs]{revtex4}

\usepackage{graphics,bm}
\usepackage{epsfig}
\usepackage{graphicx}

\newcommand{\beq}{\begin{equation}}
\newcommand{\eeq}{\end{equation}}
\newcommand{\bqa}{\begin{eqnarray}}
\newcommand{\eqa}{\end{eqnarray}}

\def\slashchar#1{\setbox0=\hbox{$#1$}                   
   \dimen0=\wd0                                         
   \setbox1=\hbox{/} \dimen1=\wd1                       
   \ifdim\dimen0>\dimen1                                
      \rlap{\hbox to \dimen0{\hfil/\hfil}}              
      #1                                                
   \else                                                
      \rlap{\hbox to \dimen1{\hfil$#1$\hfil}}           
      /                                                 
   \fi}      

\begin{document}

\parindent=10pt
\pagestyle{plain}
\font\tenrm=cmr10
\def\sumint{\hbox{$\sum$}\!\!\!\!\!\!\int}
\def\isumdiff{\hbox{${\scriptstyle \Delta}$}\hspace{-0.275cm}\int}
\def\isumint{\hbox{${\scriptstyle \Sigma}$}\hspace{-0.275cm}\int}
\def\sumdiff{\hbox{$\Delta$}\!\!\!\!\!\!\int}
\author{Jens O. Andersen} 
\email{jensoa@ntnu.no} 
\affiliation{
Department of Physics, NTNU,
Norwegian University of Science and Technology,
N-7491 Trondheim,
Norway} 

\author{Dani\"el Boer} 
\email{d.boer@few.vu.nl} 
\affiliation{
Department
of Physics and Astronomy, Vrije Universiteit, De Boelelaan 1081, 1081
HV Amsterdam, The Netherlands} 

\author{Harmen J. Warringa} 
\email{harmen@nat.vu.nl} 
\affiliation{
Department
of Physics and Astronomy, Vrije Universiteit, De Boelelaan 1081, 1081
HV Amsterdam, The Netherlands}

\title{The effects of quantum instantons on the thermodynamics of the
$\mathbb{C}P^{N-1}$ model}

\date{\today}

\begin{abstract}
  Using the $1/N$ expansion, we study the influence of quantum
  instantons on the thermodynamics of the $\mathbb{C}P^{N-1}$ model in
  1+1 dimensions. We do this by calculating the pressure to
  next-to-leading order in $1/N$, without quantum instanton
  contributions. The fact that the $\mathbb{C}P^1$ model is equivalent
  to the $\mathrm{O}(3)$ nonlinear sigma model, allows for a
  comparison to the full pressure up to $1/N^2$ corrections for $N=3$.
  Assuming validity of the $1/N$ expansion for the $\mathbb{C}P^1$
  model makes it possible to argue that the pressure for intermediate
  temperatures is dominated by the effects of quantum instantons. A
  similar conclusion can be drawn for general $N$ values by using the
  fact that the entropy should always be positive.
\end{abstract}


\maketitle

\section{Introduction}  
It was discovered by Belavin et al.~\cite{Belavin1975} that the
classical equations of motion of Euclidean QCD have 
topologically nontrivial solutions with finite action. Such instanton
solutions and the fluctuations around them contribute to physical 
quantities, as first observed by 't Hooft~\cite{tHooft1976}.
He showed that instantons give rise to an additional source of 
$\mathrm{U}(1)_A$ symmetry breaking in QCD, which e.g.
is necessary to explain the relatively large mass of the $\eta'$ meson. 
The dependence of instanton effects on 
the coupling $g$, the number of colors $N_c$ and 
the temperature $T$, has been studied extensively afterwards. 
This is usually done in the dilute instanton gas approximation, which limits 
the conclusions to weak coupling, even though the effects
of instantons are nonperturbative and typically go like
$\exp(-c/g^2)$, where $c$ is a constant. In this approximation, the
effects are exponentially suppressed in the limit $N \to
\infty$, as discussed by Witten~\cite{Witten1978}. However, if also 
instantons of large size are relevant, as e.g. for the 
topological susceptibility, then instanton contributions 
can remain in the limit $N \to \infty$. See Ref.~\cite{Schaefer2004}
for a recent discussion of instanton effects at large $N$ at $T=0$.

Instanton solutions at nonzero temperature have also been studied. 
At finite temperature, bosonic field configurations, including instantons,
have to satisfy periodic boundary conditions. 
Harrington and Shepard \cite{HS} have constructed
explicit periodic classical solutions which they called calorons. 
Gross, Pisarski and Yaffe \cite{gross1980} considered their 
effect on the partition function of QCD, including quantum corrections at the 
one-loop level. Their result applies in the weak-coupling limit, i.e.\ at 
high temperature, where the temperature provides a natural cutoff on the
instanton size and small-size instantons dominate. The effects
of instantons at low temperatures, where the coupling is large 
are very difficult to calculate. Thus studying instanton
effects in theories that are less complicated than QCD may be very useful. 

Especially field theories in 1+1 dimensions have been studied extensively 
as toy models for QCD since they share several properties.
For example, the $\mathrm{O}(N)$ nonlinear sigma model and the 
$\mathbb{C}P^{N-1}$ model are both asymptotically free theories and 
a dynamical mass is generated nonperturbatively.
The $\mathrm{O}(N)$ model has instanton solutions for $N=3$, while the 
$\mathbb{C}P^{N-1}$ admits them for all $N$.
Moreover, the $\mathrm{U}(1)$ gauge symmetry of the 
$\mathbb{C}P^{N-1}$ model generates in the large-$N$ limit 
a long-range Coulomb interaction
which in 1+1 dimension grows linearly, and hence is 
confining~\cite{DAdda1978}. This is a zero-temperature result; at
nonzero temperature the model is no longer confining~\cite{Lazarides1979}.

Instantons at finite temperature were examined in detail by
Affleck~\cite{Affleck1980a, Affleck1980b, Affleck1980c} for the
$\mathbb{C}P^{N-1}$ model. He has demonstrated, by means of the
large-$N$ expansion, that instead of the classical instanton
solutions, rather quantum instantons (quantum calorons) are of
relevance. These are stationary solutions, with quantized topological
charge, of the large-$N$ quantum effective action for the
$\mathrm{U}(1)$ gauge field.  In a low-temperature analysis, which
allows for a derivative expansion, Affleck showed that the quantum
instantons correspond to the sine-Gordon solitons, whereas at high
temperature the quantum instantons coincide with the classical
instantons (see also Ref.~\cite{Actor}).

Quantum instanton solutions may also affect thermodynamic quantities.
When one performs a $1/N$ expansion around the constant stationary
point $A_\mu =0$, one restricts to configurations with zero winding
number $Q$. In this way, one may leave out important contributions
arising from quantum instantons. In this article we will demonstrate,
by including $1/N$ corrections, that neglecting quantum instantons
even leads to unphysical results, such as a negative entropy. Since
the contributions of the configurations with nonzero winding number
are difficult to include, we will obtain in an indirect way the
combined effect of these configurations. In order to show this, we
exploit the equivalence between the $\mathbb{C}P^1$ model and the
$\mathrm{O}(3)$ nonlinear sigma model. The equivalence at the
classical level was first pointed out by
Eichenherr~\cite{Eichenherr1978}, while the quantum equivalence was
shown by Banerjee~\cite{Banerjee1994}. In Ref.\ \cite{Andersen2004} we
have obtained strong indications that the $1/N$ expansion at
next-to-leading order (NLO) yields a good approximation to the exact
pressure for the $O(N)$ nonlinear sigma models for all finite values
of $N$, down to $N=4$. For our present purposes we checked that this
also applies to $N=3$, which means that a well-behaved pressure is
obtained that differs from the $N \to \infty$ pressure by order $1/N$.
In that calculation of the pressure of the $\mathrm{O}(3)$ nonlinear
sigma model, one implicitly integrates over all (quantum) instanton
configurations. Hence, the NLO pressure should be a good approximation
to the exact pressure up to order $1/N^2$ corrections.  Because of the
equivalence to the $\mathbb{C}P^1$ model, the pressure for the latter
should be the same as that of the $\mathrm{O}(3)$ nonlinear sigma
model, upon inclusion of all quantum instantons. The difference
between the pressure of the $Q=0$ sector of the $\mathbb{C}P^1$ model
and the pressure of the $\mathrm{O}(3)$ nonlinear sigma model should
thus give the contribution to the pressure of the $\mathbb{C}P^1$
model from the topological configurations with $Q \neq 0$. Since the
$1/N$ expansions in the $\mathrm{O}(N)$ nonlinear sigma model and the
$\mathbb{C}P^{N-1}$ model are different, the NLO results of the
pressure of the $\mathrm{O}(3)$ nonlinear sigma model and that of the
$\mathbb{C}P^{1}$ model do not necessarily coincide.  But if we assume
that like in the $\mathrm{O}(N)$ nonlinear sigma model the $1/N$
expansion does not break down in the $\mathbb{C}P^{N-1}$ model for
small values of $N$ (even for $N=2$ in this case) we can make a
meaningful comparison between the pressure of the $\mathrm{O}(3)$
nonlinear sigma model and the $\mathbb{C}P^1$ model to NLO in $1/N$.
That allows us to estimate the size of the contribution of topological
contributions with $Q \neq 0$ to the pressure for $N=2$. In addition,
for general values of $N$ we can derive a lower bound on the
contribution of the topological configurations to the pressure, by
using the fact that the entropy should always be positive. In this
way, we find strong indications that the topological configurations
with $Q \neq 0$ give a large contribution to the pressure and other
thermodynamical quantities for intermediate temperatures.

The article is organized as follows. In Sec.~II the essentials of the
$\mathbb{C}P^{N-1}$ model are reviewed.  We discuss the relevant
details of the quantum effective action in Sec.~III.  The calculation
of the effective potential is explained in Sec.~IV. In Sec.~V the
results of the calculation of the pressure are presented. Finally a
summary and conclusions are given in Sec.\ VI.

\section{The $\mathbb{C}P^{N-1}$ model}
The $\mathbb{C}P^{N-1}$ model is described by the following Lagrangian
which is invariant under local $\mathrm{U}(1)$ and global
$\mathrm{SU}(N)$ transformations
\begin{equation}
  \mathcal{L} = \frac{1}{2} \partial_\mu \phi_i^* \partial^{\mu}
   \phi_i + \mathcal{L}_{\mathrm{int}}
   ,\;\;\;\; \phi_i^* \phi_i = N / {g_b^2},
      \;\;\;\;\;
  i = 1 \ldots N,
 \label{eq:cpnlagrandens}
\end{equation}
where $\phi(x)$ is a complex scalar field and $g_b$ is the bare coupling
constant. The interaction part of the Lagrangian is given by
\begin{equation}
  \mathcal{L}_{\mathrm{int}} = \frac{g_b^2}{2 N}  
   \left( \phi_i^* \partial_\mu \phi_i \right) 
  \left( \phi_j^* \partial^\mu \phi_j \right).
\label{eq:cpnint}
\end{equation}
The Lagrangian can also be written in terms of a $\mathrm{U}(1)$ gauge field
$A_{\mu}$
\begin{equation}
  \mathcal{L} = \frac{1}{2} 
  \left \vert \mathcal{D}_{\mu} \phi_i \right \vert^2
,\;\;\;\;\;\;\; \phi_i^* \phi_i = N / {g_b^2} .
  \label{eq:cpnlagrgaugefields}
\end{equation}
where $\mathcal{D}_\mu = \partial_\mu + i A_\mu$ is the covariant derivative.
By solving the equations of motion for $A_\mu$
and inserting this expression into \eqref{eq:cpnlagrgaugefields}, the
original Lagrangian Eq.~(\ref{eq:cpnlagrandens}) is recovered.

The $\mathbb{C}P^{1}$ model is equivalent to the $\mathrm{O}(3)$
nonlinear sigma model~\cite{Eichenherr1978, Banerjee1994}. 
The correspondence can be made
explicit by writing the $\mathrm{O}(3)$ nonlinear sigma fields
$\chi(x)$ as 
\begin{equation}
  \chi_a(x) = 
\sqrt{g_b^2 / N }\,
\phi^*_i(x) 
  (\sigma_a)^{\phantom{**}}_{ij} \phi^{\phantom{*}}_j(x)
  ,\;\;\;\;\;\;a = 1 \ldots 3,
  \label{eq:duality}
\end{equation}
where $\sigma_a$ are Pauli matrices. 
Using
Eq.~(\ref{eq:duality}), the Lagrangian for the $\mathrm{O}(3)$ nonlinear sigma 
model, $\mathcal{L} = \left( \partial_\mu \chi_a \right)^2 / 2$,
with the constraint $\chi_a \chi_a = N / g_b^2$ turns into the
$\mathbb{C}P^{1}$ Lagrangian, Eq.\ (\ref{eq:cpnlagrandens}), with
the corresponding constraint. 

As mentioned the $\mathbb{C}P^{N-1}$ model allows instanton solutions
for all $N$. This follows from the fact that $\mathbb{C}P^{N-1} \cong
\mathrm{SU}(N)/\mathrm{U}(N-1)$, such that $\pi_2(\mathbb{C}P^{N-1}) =
\mathbb{Z}$. For the $\mathrm{O}(N)$ nonlinear sigma models on the
other hand, the relevant coset is $\mathrm{O}(N)/\mathrm{O}(N-1) \cong
S^{N-1}$, where $\pi_2(S^{N-1}) \neq 0$ for $N=3$ only. Since one also
has a correspondence between the $\mathbb{C}P^{N-1}$ model and the
$\mathrm{O}(2N)$ nonlinear sigma model in the limit $N \to \infty$,
one can conclude that the instantons of the $\mathbb{C}P^{N-1}$ model
disappear in the limit $N\to \infty$, which coincides with the fact
that they have infinite action in this limit ($S = \pi N |Q|/g_b^2$
for the classical instantons).
 
\section{Effective action}
The constraint in Eq.~(\ref{eq:cpnlagrandens}) can be implemented by 
introducing an
auxiliary field $\alpha$. The Lagrangian Eq.~(\ref{eq:cpnlagrandens})
can then be written as
\begin{equation}
   \mathcal{L} = \frac{1}{2} 
  \left \vert \mathcal{D}_{\mu} \phi_i \right \vert^2
  - \frac{i}{2}  \alpha \left(\phi_i^* \phi^{\phantom{*}}_i
- N / g_b^2\right).
\end{equation}
Since the Lagrangian is quadratic in the $\phi$'s, they can be integrated
out exactly, resulting in the following effective action 
\begin{equation}
  S_{\mathrm{eff}} = N \mathrm{Tr}\, \ln \left (-\mathcal{D}_{\mu} 
    \mathcal{D}^{\mu} - i \alpha \right)  
 + i \frac{N}{2g_b^2} \int_{X}\alpha(x),
   \label{eq:effectiveactionCPN}
\end{equation}
where the subscript $X$ indicates integration over two-dimensional
Euclidean space. The vacuum expectation value of
the $\alpha$ field is purely imaginary and can therefore be written as
$\alpha= i m^2 + \tilde \alpha/\sqrt{N}$, where $\langle
\alpha\rangle=im^2$ and $\tilde{\alpha}$ a quantum fluctuating field.
The scaling of the quantum fluctuating field with a factor
$1/\sqrt{N}$ is merely a convenient way of implementing the $1/N$
expansion and has no effect on the final results. This yields
\begin{multline}
  S_{\mathrm{eff}} = N \mathrm{Tr} \ln \left [-\partial^2 + m^2 - i 
 \left \{ \partial_{\mu}, A^{\mu} \right \}
  + A_{\mu} A^{\mu} \right.
\\
\left.
- i
  \frac{\tilde \alpha}{\sqrt{N}}
   \right]
  - \frac{N}{2 g_b^2} \int_{X} 
  \left[m^2 - \frac{i}{\sqrt{N}} \tilde \alpha(x) \right]. 
 \label{eq:effaction} 
\end{multline}
Affleck~\cite{Affleck1980a} showed that $S_\mathrm{eff}$ has 
stationary solutions $A_\mu$ at finite temperature that have a
quantized topological charge. Since these solutions are stationary
points of an action in which quantum effects are incorporated, they
are called `quantum instantons'. Such instantons need to be considered 
in a full calculation of the pressure. As a first step to investigate the
relevance of the quantum instantons, we take 
into account fluctuations around the trivial vacuum $A_\mu = 0$. We will
do this in a way consistent with the $1/N$ expansion by 
scaling the gauge fields with a factor $1/\sqrt{N}$ as well. In the 
$1/N$ expansion around the stationary point $A_\mu=0$, quantum
instantons do not arise, as their nonzero boundary values (in the $A_1=0$
gauge, these are $A_0(x_0,\pm \infty) = 2 \pi n_\pm T$, where $Q=n_+-n_-$)   
will not be achieved~\cite{Affleck1980a}. 
To next-to-leading (NLO) order in $1/N$, one obtains~\cite{DAdda1978}
\bqa\nonumber
  S_{\mathrm{eff}} &=& N \mathrm{Tr} \ln  \left (-\partial^2 + m^2 \right)
  -\frac{N m^2}{2 g^2_b} \beta V \\ \nonumber
&&
  + i \frac{\sqrt{N}}{2} 
  \int_{X} \tilde \alpha (x) 
\left ( \frac{1}{g^2_b} -\int_P \frac{1}{P^2 + m^2} \right)
\nonumber  \\ && \nonumber
+ \frac{1}{2} \int_{X,Y} 
\tilde \alpha(x) \Gamma(x - y) \tilde \alpha(y)
 \\ &&
 + \frac{1}{2} \int_{X,Y}A^{\mu}(x) \Delta_{\mu \nu}(x - y)  A^{\nu}(y),
 \label{eq:effaction2}
\eqa
where $\beta = 1/T$ is equal to the inverse temperature
and $V$ is the volume of the one-dimensional space.
Equation (\ref{eq:effaction2}) shows that although a kinetic term for
the gauge fields is absent in the classical action, 
such a term is generated by quantum fluctuations. 
Its tensorial structure at finite temperature is the same
as at zero temperature, which is specific to $1+1$ dimensions.
One finds~\cite{DAdda1978, Lazarides1979}
\bqa\nonumber 
\Gamma(P) &=&
  \frac{1}{2 \pi} \frac{1}{\sqrt{P^2 (P^2 + 4m^2)}} 
\\ 
&&
\times
\ln \left(
  \frac{\sqrt{P^2 + 4 m^2} + \sqrt{P^2}}{\sqrt{P^2 + 4m^2} - \sqrt{P^2}}
  \right)+ \Gamma_T(P) ,
\\ 
\nonumber
\\
 \Delta_{\mu \nu}(P) &=&
  \left(\delta_{\mu \nu} - \frac{P_{\mu} P_{\nu}}{P^2} \right) 
  \Delta_{\mu}^{\mu}(P),
\eqa
where $P=(p,p_0)$ and 
\bqa\nonumber
\Delta_{\mu}^{\mu}(P) &=& \\ \nonumber
&&  
\hspace{-2cm}
\frac{1}{2 \pi} \left[\sqrt{\frac{P^2 + 4m^2}{P^2}} \ln \left(
  \frac{\sqrt{P^2 + 4 m^2} + \sqrt{P^2}}{\sqrt{P^2 + 4m^2} - \sqrt{P^2}}\right)
 - 2
  \right] \\
&& + (P^2 + 4m^2) \Gamma_T(P),
\\
 \Gamma_T(P) &=& 
\frac{1}{\pi} \int_{-\infty}^{\infty}\frac{\mathrm{d} q}{\omega_q} 
\frac{P^2 + 2pq}{ (P^2 + 2pq)^2 + 4 p_0^2 \omega_q^2}
n(\omega_q).
\eqa
Here $n(\omega_q) = (\exp(\beta \omega_q) -1)^{-1}$ is the Bose-Einstein 
distribution function and $\omega_q = \sqrt{q^2+m^2}$.

\section{Effective potential}
\label{ep}
The leading-order (LO) contribution to the effective potential can be
read off directly from Eq.~(\ref{eq:effaction2}).  The next-to-leading
order corrections, is obtained by carrying out a Gaussian integration
over the fluctuations $\tilde \alpha$ and $A_\mu$.  In order to do so,
one has to fix a gauge, and throughout the paper we employ the
generalized Lorentz gauge.  Including the contribution from the ghost,
we obtain the contribution to the effective potential from the gauge
field:
\bqa\nonumber
  \mathcal{V}_{\mathrm{gauge}}(m^2) 
&=&
\sumint_P \ln P^2 - \frac{1}{2} \sumint_P \ln \mathrm{det}\,
 \left ( \Delta_{\mu \nu} + \frac{1}{\xi} P_\mu P_\nu \right) \\
&=  & 
\frac{1}{2} \sumint_P \ln P^2 
-\frac{1}{2} \sumint_P \ln \Delta_{\mu}^{\mu},
\label{gaugeind}
\eqa
where the sum-integral is defined as
\bqa
\sumint_P&=&\sum_{p_0=2\pi n T}\int 
\frac{\mathrm{d}p}{2\pi}\:.
\eqa
We emphasize that Eq.~(\ref{gaugeind}) is independent of the gauge-fixing
condition.
From Eq.\ (\ref{eq:effaction2}) and the results above, we obtain the
following finite temperature effective potential up to next-to-leading
order in $1/N$.
\begin{equation}
 \mathcal{V}(m^2) = N \mathcal{V}_\mathrm{LO}(m^2) +
 \mathcal{V}_\mathrm{NLO}(m^2) ,
\end{equation}
where 
\bqa
 \mathcal{V}_\mathrm{LO}(m^2) &=& 
  \frac{m^2}{2 g_b^2} - \sumint_P \ln(P^2 + m^2), \\ \nonumber
 \mathcal{V}_\mathrm{NLO}(m^2) &=&
  - \frac{1}{2} \sumint_P \ln \Gamma(P)
  - \frac{1}{2} \sumint_P \ln \Delta_{\mu}^{\mu}(P)
 \\
&&
  + \frac{1}{2} \sumint_P \ln P^2.
\label{uv}
\eqa
The effective potential is ultraviolet divergent. To regulate
the divergences, we introduce an ultraviolet momentum cutoff $\Lambda$.
After subtracting $T$ and $m$-independent infinite constants, we obtain
\begin{equation}
 \mathcal{V}_\mathrm{LO}(m^2) =
  \frac{m^2}{2 g_b^2} - \frac{m^2}{4\pi} 
   \left[1 + \ln \left(\frac{\Lambda^2}{m^2} \right) \right]
     + \frac{1}{4\pi} T^2 J_0 (\beta m),
\end{equation}
where
\begin{equation}
  J_0(\beta m) = \frac{8}{T^2} \int_0^\infty \frac{\mathrm{d}q\,
  q^2}{\omega_q} n(\omega_q).
\end{equation}
The minimum of the leading-order effective potential obeys
the following gap equation
\begin{equation}
  \frac{1}{g_b^2} = \frac{1}{2 \pi} \ln 
\left( \frac{\Lambda^2}{m^2} \right)
  + \frac{1}{2\pi} J_1(\beta m),
 \label{eq:logapeqcpn}
\end{equation}
where
\begin{equation}
J_1({\beta m}) = 4 \int_0^\infty \frac{\mathrm{d}q\,}{\omega_q}\, n(\omega_q).
\end{equation}
In order to calculate the NLO order contribution
to the effective potential, we write $\mathcal{V}_\mathrm{NLO}$ 
as a sum of divergent ($D$) and finite parts ($F$) in
the following way
\begin{multline}
 \mathcal{V}_\mathrm{NLO}(m^2) = 
     - \frac{1}{2} \left( D_1 + D_2 + F_1 + F_2 + F_3 + F_4 \right)
\\ - \frac{\pi}{3}T^2,
  \label{eq:effpot1}
\end{multline}
where the divergent and finite quantities are defined through the
following relations
\begin{equation}
\begin{split}
   D_1 + F_1 = \int_P \ln \tilde \Gamma(P) , & \;\;\;\;\;\;
    F_3 = \sumdiff_P \ln \tilde \Gamma(P) 
   , \\
 D_2 + F_2 = \int_P \ln \tilde \Delta_{\mu}^{\mu}(P)
 , & \;\;\;\;\;\;
       F_4 = \sumdiff_P \ln \tilde \Delta_{\mu}^{\mu}(P).
\end{split}
\end{equation} 
Here $\tilde \Gamma(P) \equiv 2 \pi \sqrt{P^2(P^2 + 4m^2)} \Gamma(P)$,
$\tilde\Delta^{\mu}_{\mu}(P) \equiv 2\pi \sqrt{P^2 / (P^2 + 4m^2)}
\Delta_{\mu}^{\mu}(P)$, and
\bqa
\sumdiff_P &\equiv&\sumint_P-\int_P.
\eqa
The functions $D_1$ and $D_2$ contain the ultraviolet 
divergences of the NLO order effective potential. In order
to isolate these divergences, the high-momentum limit of
$\tilde \Gamma(P)$ and $\tilde \Delta_\mu^\mu(P)$ are needed.
In the high-momentum approximation ($|p| \gg T$), 
we obtain
\begin{eqnarray}
  \tilde \Gamma(P)
   &\approx& 
   \ln\left(\frac{P^2}{\bar m^2} \right) + \frac{2m^2}{P^2}
        + \frac{2 m^2 J_1(\beta m)}{P^2}\left(1 - \frac{2 p_0^2}{P^2} \right)
, 
\nonumber \\
&&
 \label{eq:gammaExp}
 \\
\tilde \Delta_{\mu}^{\mu}(P)
 &\approx& 
\ln\left(\frac{P^2}{\bar m_e^2} \right) + \frac{6m^2}{P^2}
 + \frac{2 m^2 J_1 (\beta m)}{P^2} \left(1 - \frac{2 p_0^2}{P^2} \right),
 \nonumber \\ &&       
  \label{eq:deltaExp}
\end{eqnarray}
where $\bar m^2 = m^2 \exp[- J_1(\beta m) ]$ and $\bar m^2_e = m^2
\exp[2 - J_1(\beta m)]$.
The divergences $D_1$ and $D_2$ can be obtained by integrating
\eqref{eq:gammaExp} and
\eqref{eq:deltaExp} over spatial momenta and we find
\begin{eqnarray}
  D_1 &=& \frac{1}{4\pi} \left[
  \Lambda^2 \ln \ln \left( \frac{\Lambda^2}{\bar m^2} \right)
  - \bar m^2 \mathrm{li}\, \left( \frac{\Lambda^2}{\bar m^2} \right)
  \right. \nonumber \\
& & 
\left. + 2 m^2 \ln \ln \left( \frac{\Lambda^2}{\bar m^2} \right)
      \right ],
\\ 
   D_2 &=& \frac{1}{4\pi} \left[
     \Lambda^2 \ln \ln \left( \frac{\Lambda^2}{\bar m_e^2} \right)
 - \bar m_e^2 \mathrm{li}\, \left( \frac{\Lambda^2}{\bar m_e^2} \right)
\right.
 \nonumber \\
& & \left. 
    + 6 m^2 \ln \ln \left( \frac{\Lambda^2}{\bar m_e^2} \right)
\right],
\end{eqnarray}
where $\mathrm{li}(x)$ is the logarithmic integral defined by
\begin{equation}
\mathrm{li}(x) = \mathcal{P} \int_0^x \mathrm{d}t \, \frac{1}{\ln t}.
\end{equation}
Here ${\cal P}$ denotes the principal-value prescription.
From $D_1$ and $D_2$ it can be seen that (through the dependence on
$\bar m^2$ and $\bar m_e^2$) the effective potential contains
temperature-dependent divergences. They cannot be eliminated in a
temperature-independent way. See Ref.~\cite{harmenproc} for a detailed
discussion of the occurrence of these divergences.
However, they become
temperature-independent at the minimum of the effective
potential (see Sec.\ V). 

\begin{figure}[htb]
\begin{center}
\includegraphics{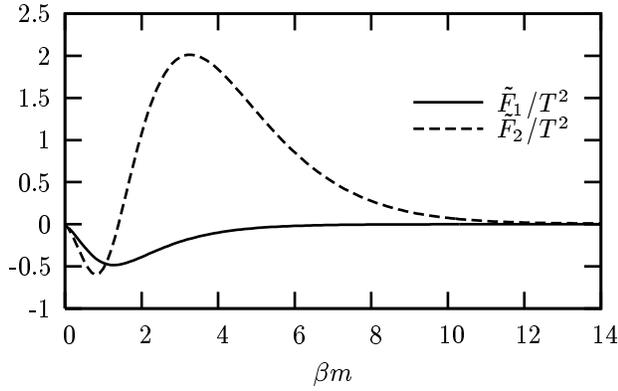}
\caption{$\tilde F_1(\beta m) /T^2$ and 
$\tilde F_2(\beta m)/T^2$ plotted as functions of $\beta$m.} 
\label{fig:f1f2}
\end{center}
\end{figure}

The finite functions $F_1$ and $F_2$ will be obtained numerically. In
order to calculate these functions, we write the divergences 
partly in terms of an integral. This prevents subtracting large
quantities which can give rise to big numerical errors. The functions
$F_1$ and $F_2$ are calculated using the following expressions
\begin{eqnarray}
  F_1\!\! &=& \!\!\mathcal{P} \int_P \ln \left [\frac{\tilde \Gamma(P)}{\ln
       \left(P^2 / \bar m^2\right)} \right] - 
    \frac{2 m^2}{4\pi} \ln \ln \left(
       \frac{\Lambda^2}{\bar m^2} \right),
   \\
  F_2\!\! &=& \!\!\mathcal{P}
       \int_P \ln
 \left[
   \frac{\tilde \Delta_\mu^\mu(P)}{\ln \left(P^2 / \bar m_e^2\right)}
\right]
 - \frac{6 m^2}{4\pi} \ln \ln \left( \frac{\Lambda^2}{\bar m_e^2} \right).
\end{eqnarray}
At zero
temperature it was found that $F_1 \approx m^2 \gamma_E / (2 \pi)$
and $F_2 \approx m^2 c_1 / (2 \pi)$, where $c_1 \approx 0.611 671
457 \ldots$ ~\cite{Campostrini1992}. 
For convenience the finite-temperature 
parts of $F_1$ and $F_2$ are defined as $\tilde F_{1} =
F_1 - m^2 \gamma_E / (2\pi)$ and $\tilde F_{2} = F_2 - m^2 c_1 /
(2\pi)$. 
These functions divided by $T^2$ depend on $\beta m$ only and are
displayed in Fig.~\ref{fig:f1f2}. 
In the limit $\beta m\rightarrow\infty$,
these functions go to zero because 
the temperature-dependent parts of the inverse propagators are exponentially
suppressed compared to the zero-temperature contribution.
For small $\beta m$, these functions also go to zero as can be inferred from
the limit $\beta m\rightarrow0$ of $\Gamma(P)$. This limit can be found by
first performing a momentum integration and then noting that the 
dominant contribution arises from the zeroth Matsubara mode. This yields
\begin{equation}
  \Gamma(P) \approx \frac{1}{\beta m} \frac{P^2}{P^4 + 4m^2 p^2}.
\label{eq:lowbetamappr}
\end{equation}

\begin{figure}[htb]
\begin{center}
\includegraphics{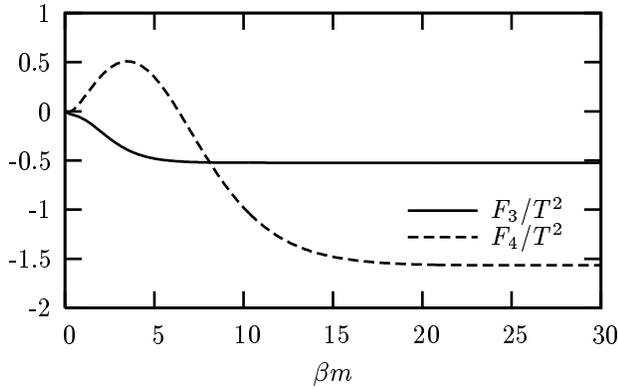}
\caption{$F_3(\beta m)/T^2$ and $F_4(\beta m)/T^2$
as functions of $\beta m$.} 
\label{fig:f3f4}
\end{center}
\end{figure}

The finite functions $F_3/T^2$ and $F_4/T^2$ were calculated using a
modified Abel-Plana formula \cite{Andersen2004b}. They are displayed
in Fig.~\ref{fig:f3f4}.

The large-$\beta m$ limit of $F_3$ and $F_4$ can be obtained by noting
that for large $\beta m$, the temperature-dependent part of the
inverse propagator does not contribute to $F_3$ and $F_4$.
Furthermore, the dominant contribution to the difference of a
sum-integral and an integral arises from the low-momentum modes. Thus
the large-$m$ behavior of the zero-temperature inverse propagators can
be used to obtain a large-$\beta m$ approximation for $F_3$ and $F_4$:
\begin{equation} 
F_3 \approx \frac{1}{2} \sumdiff_P \ln P^2 = - \frac{\pi}{6} T^2,
\;\;\;F_4 \approx \frac{3}{2} \sumdiff_P \ln P^2 
= - \frac{\pi}{2} T^2.
\end{equation}
As can be seen in Fig.\ \ref{fig:f3f4} this is in agreement with the
numerical calculations.
The small-$\beta m$ limit of $F_3/T^2$ and $F_4/T^2$ can be obtained too, 
using the small-$\beta m$ limit of $\Gamma(P)$. 
The result $F_3 \approx 0$ and $F_4 \approx 0$ is in
agreement with the numerical calculations 
displayed in Fig.~\ref{fig:f3f4}.

\section{Contribution of quantum instantons to the pressure}
\begin{figure}[htb]
\includegraphics{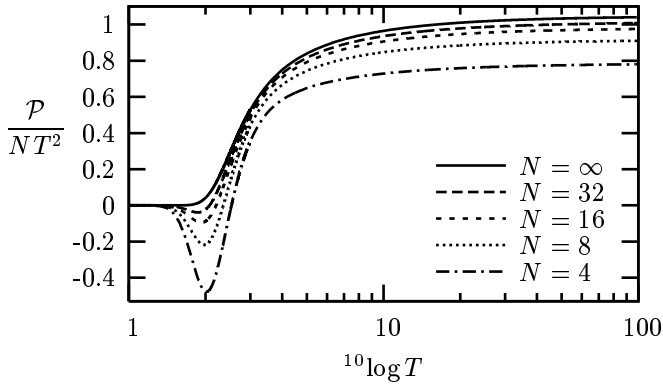}
\caption{Contribution from the zero-winding-number configurations to the 
pressure $\mathcal{P}$ of the $\mathbb{C}P^{N-1}$ model normalized to
$NT^2$, as a function of temperature and different values of $N$.}
\label{fig:cpnpressure}
\end{figure}

\begin{figure}[htb]
\includegraphics{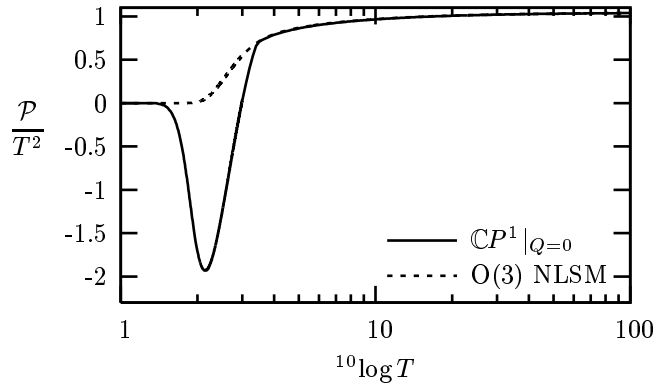}
\caption{The normalized pressure $\mathcal{P}/T^2$ of the $\mathrm{O}(N)$ 
nonlinear sigma model to NLO in $1/N$ for $N=3$ compared to the
contribution to the pressure of the $\mathbb{C}P^{1}$ model
from the configurations with zero winding number.}
\label{fig:cp1o3pres}
\end{figure}

\begin{figure}[htb]
\includegraphics{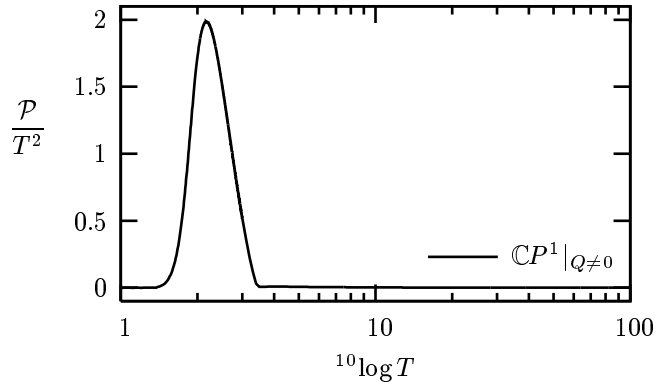}
\caption{Estimate of the contribution from 
the configurations with $Q\neq0$ to the 
normalized pressure $\mathcal{P}/T^2$ of 
the $\mathbb{C}P^{1}$ model.}
\label{fig:cpnpresinstcontr}
\end{figure}

In Sec.~\ref{ep}, the effective potential was evaluated and it
was found that it contains temperature-dependent
ultraviolet divergences. At the minimum these temperature-dependent
divergences will disappear as will be discussed now. To calculate the
effective potential at the minimum, one only needs to solve the leading-order
gap equation (\ref{eq:logapeqcpn}) as was shown by
Root~\cite{Root1974}. 
As a result, the LO and NLO order contributions to
the pressure are given by
\begin{eqnarray}
  \mathcal{P}_{\mathrm{LO}} 
&=& 
     \mathcal{V}^{T}_{\mathrm{LO}}(m^2_T) 
   - \mathcal{V}^{T=0}_{\mathrm{LO}}(m^2_0), 
\\
  \mathcal{P}_{\mathrm{NLO}} &=&
     \mathcal{V}^{T}_{\mathrm{NLO}}(m^2_T) 
   - \mathcal{V}^{T=0}_{\mathrm{NLO}}(m^2_0)\; ,
\end{eqnarray}
where $m^2_T$ is the solution of the leading-order gap equation
(\ref{eq:logapeqcpn}) at temperature $T$. By using the leading-order gap
equation, it can be shown that at the minimum
the divergent terms $D_1$ and $D_2$ become
\begin{eqnarray}
 D_1 &=& \frac{\Lambda^2}{4\pi} \left[
 \ln \left( \frac{2\pi}{g_b^2} \right)
 - \exp \left( - \frac{2 \pi}{g_b^2} \right) 
  \mathrm{li} \exp \left( \frac{2 \pi}{g_b^2} \right)
 \right]
\nonumber\\
&&
  + \frac{2 m_T^2}{4\pi} 
\ln \ln \left( \frac{\Lambda^2}{\bar m_T^2} \right), \\
  D_2&=& 
\frac{\Lambda^2}{4\pi} \left[
 \ln \left( \frac{2\pi}{g_b^2}{-}2 \right)
 - \exp \left(2{-} \frac{2 \pi}{g_b^2} \right) 
  \mathrm{li} \exp \left( \frac{2 \pi}{g_b^2} {-}2\right)
 \right]
\nonumber\\
&&
  + \frac{6 m_T^2}{4\pi} 
\ln \ln \left( \frac{\Lambda^2}{\bar m_{eT}^2} \right).
\end{eqnarray}
Hence, the temperature-dependent quadratic divergence and the
divergence proportional to li(x) become
temperature-independent at the minimum of the effective
potential. 
As a result these divergences can be eliminated by counterterms 
that are independent of temperature. Furthermore, the divergences
proportional to $\ln \ln$ can be eliminated by the 
coupling-constant renormalization, which amounts to the substitution
$g_b^2\rightarrow Z_g^2g^2$, where
\begin{equation}
\frac{1}{Z_g^2} = 1+ \frac{g^2}{2\pi} 
 \ln \left ( \frac{\Lambda^2}{\mu^2} \right) + \frac{2}{N}
  \frac{g^2}{\pi} \ln \ln \left( \frac{\Lambda^2}{\mu^2}\right).
\end{equation}
Using the results above, it follows that the leading and
next-to-leading order contributions to the pressure are given by
\bqa\nonumber
  \mathcal{P}_\mathrm{LO} &=& \frac{m^2_T}{2 g^2} - \frac{m^2_T}{4\pi}
 \left[1 + \ln \left( \frac{\mu^2}{m_T^2} \right) \right] +
 \frac{T^2}{4\pi} J_0(\beta m_T) 
\\ &&
+ \frac{m^2_0}{4\pi},
\\ \nonumber
 \mathcal{P}_\mathrm{NLO} &=&
- \frac{1}{2} \left[ \tilde F_1(m_T) + \tilde F_2(m_T) + F_3(m_T) +
F_4(m_T) \right]
\\
&&
 + \frac{1}{4\pi} (\gamma_E + c_1) (m_0^2 - m_T^2) 
   - \frac{\pi}{3}T^2 .
\label{eq:cpnpress}
\eqa
The results of the calculation of the pressure are displayed in
Fig.~\ref{fig:cpnpressure} for the arbitrary choice $g^2(\mu = 500) =
10$ and different values of $N$. As one can see, for low temperatures
and all finite values of $N$ the pressure first decreases with
increasing $T$. A decreasing pressure implies that the entropy becomes
negative. Clearly, this is in conflict with the third law of
thermodynamics, which states that the entropy is minimal at zero
temperature. As we will remind the reader below, we have strong
indications that the $1/N$ expansion itself is not the reason for the
negative pressure, therefore, it is likely the fact that the effective
action, Eq.~(\ref{eq:effaction}), was expanded only around the vacuum
$A_\mu = 0$ solution with zero winding number. The contributions from
the other vacua with nonzero winding number were left out in the
calculations. As one can see from the figure, the problem of the
negative pressure becomes less severe as $N$ becomes larger. This is
in agreement with the fact that the instanton contribution vanishes in
the limit $N \rightarrow \infty$.  Moreover, the problem disappears as
the temperature increases, which is consistent with the fact that at
weak coupling the effects of instantons become highly suppressed as
function of $T$~\cite{gross1980}.
Using the equivalence with the $\mathrm{O}(3)$ nonlinear sigma model,
the contribution from the quantum instantons with nonzero winding
number to the pressure of the $\mathbb{C}P^{1}$ model can be
estimated.  Because the integration over all scalar-field
configurations is done exactly in the $\mathrm{O}(3)$ nonlinear sigma
model, including those with $Q\neq0$, the effects of all quantum
instantons are automatically included in the large-$N$ quantum
effective potential for $\tilde{\alpha}$. In Fig.~\ref{fig:cp1o3pres},
the result of the NLO order calculation of the pressure of the
$\mathrm{O}(N)$ nonlinear sigma model for $N=3$ is compared to the
contribution of the configurations with $Q=0$ to the pressure of the
$\mathbb{C}P^{1}$ model.  Fig.~\ref{fig:cp1o3pres} shows that for very
low and high temperatures the pressures coincide, while for
intermediate temperatures they are very different. This difference is
displayed in Fig.~\ref{fig:cpnpresinstcontr} and is a strong hint that
quantum instantons give a sizable contribution to the pressure in the
region where the pressure increases considerably.

As we have already mentioned in the introduction, in Ref.\
\cite{Andersen2004} we have obtained strong indications that the $1/N$
expansion yields trustworthy results for the $O(N)$ nonlinear sigma
models for all finite values of $N$, down to $N=4$. The $N=3$ pressure
presented here is in full agreement with the results obtained earlier
and there is no reason to believe that the $1/N$ expansion for $N=3$
is not to be trusted. The NLO corrections for the $\mathrm{O}(N)$
nonlinear sigma model are of the expected order $1/N$.  The pressure
of $\mathrm{O}(3)$ nonlinear sigma model evaluated to NLO in the $1/N$
expansion includes the effects of all quantum instantons (up to
$1/N^2$ corrections).  Therefore, we believe that we have obtained a
good approximation to the exact $\mathbb{C}P^1$ model pressure. As
said, this is up to order $1/N^2$ corrections, which cannot solve the
discrepancy with the pressure of the $Q=0$ sector of the
$\mathbb{C}P^1$ model.

Since the entropy has to be positive, for general values of $N$ we can
estimate a lower bound on the contribution of the quantum instantons
with $Q\neq 0$ to the pressure.  The lower bounds turn out to be
almost independent of $N$ and have a similar shape with a somewhat
lower maximum than the estimated contribution for $N=3$, displayed in
Fig.\ \ref{fig:cpnpresinstcontr}.

\section{Summary and Conclusions}
In this article the effect of quantum instantons on thermodynamical
quantities of the $\mathbb{C}P^{N-1}$ model was investigated. We
expanded the effective potential of the $\mathbb{C}P^{N-1}$ model
around the trivial vacuum, and calculated it to NLO order in $1/N$. It
was shown that the effective potential contains temperature-dependent
divergences which only can be renormalized at the minimum of the
effective potential. Hence thermodynamic quantities can be rendered
finite as in the (non)linear sigma model~\cite{Andersen2004,
  Andersen2004b}.

We found that for finite $N$, the contribution from the vacuum with
$Q=0$ gives rise to a negative pressure for intermediate temperatures
where the leading-order pressure increases rapidly. Since this is
unphysical, it indicates that quantum instantons contribute
significantly to the pressure in this temperature range. In agreement
with the disappearance of the instantons in the limit $N \rightarrow
\infty$, the problem of the negative pressure becomes less severe for
large values of $N$.

For the $\mathbb{C}P^1$ model, we found the contribution of the
quantum instantons by using its (quantum) equivalence to the
$\mathrm{O}(3)$ nonlinear sigma model. In the $1/N$ approximation to
the $\mathrm{O}(3)$ nonlinear sigma model, one implicitly integrates
over all quantum instantons and finds a well-behaved, increasing
pressure at next-to-leading order in $1/N$ that should be a good
approximation to the exact pressure, up to $1/N^2$ corrections, even
for $N=3$~\cite{Andersen2004}.  Assuming that the $1/N$ expansion does
not break down in the $\mathbb{C}P^{N-1}$ model for small values of
$N$ as well, we have compared the next-to-leading order calculation of
the pressure of the $\mathbb{C}P^1$ model without quantum instantons
to the pressure of the $\mathrm{O}(3)$ nonlinear sigma model. This
comparison allowed us to estimate the contribution of quantum
instantons with non-zero winding number to the pressure for $N=2$.
For general values of $N$ we were able to estimate a lower bound on
the contribution of the quantum instantons by using the fact that the
entropy should be positive.  Calculating explicitly the contribution
from quantum instantons with winding number $Q=1$ would be a useful
step to get a deeper understanding of this issue.

\end{document}